# O-Band Subwavelength Grating Filters in a Monolithic Photonics Technology

Francis O. Afzal, Yusheng Bian, Bo Peng, Shuren Hu, Kevin Dezfulian, Karen Nummy, Andy Stricker, Abdelsalam Aboketaf, Crystal Hedges, Zoey Sowinski, Michal Rakowski, Won Suk Lee, Rod Augur, Dave Riggs, Ken Giewont, and Sharon M. Weiss

*Abstract*—The data communications industry has begun transitioning from electrical to optical interconnects in datacenters in order to overcome performance bottlenecks and meet consumer needs. To mitigate the costs associated with this change and achieve performance for 5G and beyond, it is crucial to explore advanced photonic devices that can enable high-bandwidth interconnects via wavelength-division multiplexing (WDM) in photonic integrated circuits. Subwavelength grating (SWG) filters have shown great promise for WDM applications. However, the small feature sizes necessary to implement these structures have prohibited them from penetrating into industrial applications. To explore the manufacturability and performance of SWG filters in an industrial setting, we fabricate and characterize O-band subwavelength grating filters using the monolithic photonics technology at GLOBALFOUNDRIES (GF). We demonstrate a low drop channel loss of -1.2 dB with a flat-top response, a high extinction ratio of -30 dB, a 3 dB channel width of 5 nm and single-source thermal tunability without shape distortion. This filter structure was designed using elements from the product design kit provided by GF and functions in a compact footprint of 0.002 mm$^2$ with a minimum feature size of 150 nm.

## I. INTRODUCTION

As global datacom performance demands continue to increase, commercial and academic sectors continue to investigate how communications technology can evolve to satisfy growing consumer needs. One of the greatest factors currently limiting datacom bandwidth and energy efficiency lies in short-range (< 2 km) interconnects which transmit data inside datacenters. The recent trend to combat this bottleneck has been for companies to switch from electrical to optical interconnects, as optical interconnects are able to simultaneously reduce parasitic losses and increase bandwidth compared to their electrical counterparts [1].

The bandwidth increase for optical interconnects is enabled in large part by wavelength-division multiplexing (WDM). In WDM, various data channels are encoded onto separate wavelengths of light which can propagate together in a single waveguide or fiber and can be routed to other waveguides and fibers. By combining multiple wavelength channels of light into a single optical interconnect, the interconnect bandwidth can be multiplied by the number of wavelength channels used. Since WDM techniques can be used in addition to pulse-amplitude modulation (PAM) and quadrature amplitude modulation (QAM), enormous data rates > 1Tb/s could potentially be realized for individual optical interconnects [2].

Despite large improvements to power consumption, bandwidth and cost scaling of optical interconnects over their electrical counterparts, transitioning to optical interconnects incurs an implementation cost of purchasing optical transceivers that process the electro-optical signal conversion in the form of a discrete device [2]. To reduce the cost of optical transceivers, GLOBALFOUNDRIES (GF) have pioneered monolithic silicon photonics technologies that enable the large-scale fabrication of photonics and CMOS technologies on the same chip [3].

To realize WDM functionality in photonic integrated circuits (PICs) on a CMOS technology platform, many photonic filtering approaches have been explored, including arrayed waveguide gratings (AWGs), Echelle gratings, cascaded Mach-Zehnder interferometers (MZIs) and coupled rings [4]–[6]. While these devices typically work on feature scales readily compatible with commercial photolithography, demonstrated performance in the categories of channel width, insertion loss, channel shape and active tunability either make them unsuitable for many commercial applications or leave significant room for improving performance or implementation. Recent work on add/drop subwavelength grating (SWG) filters has shown great promise for achieving low losses, low crosstalk, flat-top channel shapes and channel widths spanning CWDM and DWDM requirements in a serially cascadable and modular

This work was supported in part by the National Science Foundation under grant ECCS1809937 (GOALI).

Yusheng Bian, Bo Peng, Shuren Hu, Kevin Dezfulian, Karen Nummy, Andy Stricker, Abdelsalam Aboketaf, Crystal Hedges, Zoey Sowinski, Michal Rakowski, Won Suk Lee, Rod Augur, Dave Riggs and Ken Giewont, are with GlobalFoundries, Hopewell Junction, NY 12533 USA (e-mail: yusheng.bian@globalfoundries.com; bo.peng@globalfoundries.com; shuren.hu@globalfoundries.com, kevin.dezfulian@globalfoundries.com; karen.nummy@globalfoundries.com; andy.stricker@globalfoundries.com; abdelsalam.aboketaf@globalfoundries.com; crystal.hedges@globalfoundries.com; zoey.sowinski@globalfoundries.com; michal.rakowski@globalfoundries.com; wonsuk.lee@globalfoundries.com; rod.augur@globalfoundries.com; dave.riggs@globalfoundries.com; ken.giewont@globalfoundries.com ).

F. O. Afzal and S. M. Weiss are with the Department of Electrical Engineering and Computer Science, Vanderbilt University, Nashville, TN 37235 USA (e-mail: francis.afzal@vanderbilt.edu; sharon.weiss@vanderbilt.edu).

platform [7]–[10]. However, the small feature sizes of SWG filters are typically realized using electron beam lithography, which is incompatible with commercial fabrication. Pushing UV lithography in CMOS foundries to resolve such features has been explored to overcome this hurdle [10]–[13]. Moreover, most prior work with add/drop SWG filters has been conducted in the C-band [7]–[10] while the industry standard for datacenter applications is O-band operation [2]. Investigating the performance and manufacturability of SWG filters in the O-band is critical for exploring their potential to advance integrated photonics in data communications [12]. Towards this goal, we demonstrate the first add/drop SWG filters in the O-band manufactured in a CMOS foundry using a tape-out on the 90nm, monolithic silicon photonics technology at GF.

## II. DEVICE OPERATION CONCEPT

Many SWG filters are designed to operate based on contra-directional reflection [10]. This reflection is generally induced by a periodic corrugation (a general form of a grating) acting on the field of a waveguide mode. In the configuration reported here (Fig. 1), a SWG acts on the evanescent field of a bus waveguide mode and, for one frequency band, couples the waveguide mode to the backward propagating SWG mode. An intuitive explanation of this effect is that for one frequency band, first order grating diffraction provides sufficient momentum change in the waveguide mode to evanescently couple it into the backwards propagating SWG mode.

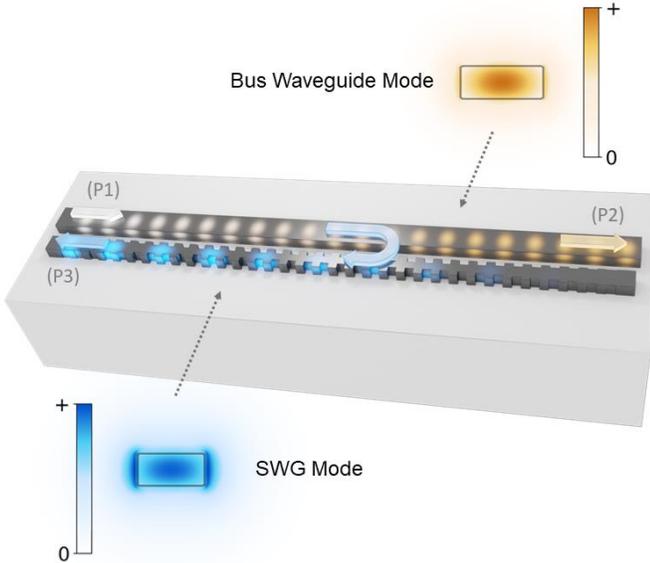

Fig. 1. Illustration of operation of SWG filter. P1, P2 and P3 refer to the input, through, and drop ports, respectively. The SWG filter is designed such that a select band of frequencies (shown in blue) is dropped into P3 while the remainder of the spectrum passes from P1 to P2. The drop frequency depends on the effective indices of the SWG and bus waveguide modes.

More formally, as the evanescent field of the waveguide mode interacts with the grating, the propagation vector of the mode, $\beta_{wvg}$, changes according to the grating diffraction relation:

$$\beta_m = \beta_{wvg} + \frac{2\pi \cdot m}{\Lambda}, \qquad (1)$$

where $\Lambda$ is the pitch of the grating (shown in Fig. 2), $m$ is an integer (either positive or negative) and $\beta_m$ is the resulting wave-vector from m$^{th}$ order diffraction. Given a SWG guided mode with propagation vector $\beta_{SWG}$, it is possible to contra-directionally diffract the evanescently coupled waveguide mode, with propagation vector $\beta_{wvg}$, into the SWG by choosing $m = -1$ in (1) and setting $\beta_{-1} = -\beta_{SWG}$, resulting in the relation

$$-\beta_{SWG} = \beta_{wvg} - \frac{2\pi}{\Lambda}. \qquad (2)$$

Placing (2) in terms of effective index and the dropped wavelength, $\lambda_{drop}$, then gives

$$-\frac{n_{SWG}}{\lambda_{drop}} = \frac{n_{wvg}}{\lambda_{drop}} - \frac{1}{\Lambda}, \qquad (3)$$

where $n_{SWG}$ and $n_{wvg}$ are the effective indices of the SWG mode and waveguide modes at $\lambda_{drop}$, respectively. By re-arranging (3) to solve for $\lambda_{drop}$, we arrive at a convenient expression for the contra-directional coupling condition.

$$\lambda_{drop} = \Lambda(n_{wvg} + n_{SWG}) \qquad (4)$$

At this point, it is important to note that a SWG can enable diffraction between any two available modes that satisfy (1) in its proximity. Back-reflections occur in the waveguide at $\lambda = \lambda_{b.r.1}$ and SWG at $\lambda = \lambda_{b.r.2}$, where:

$$\lambda_{b.r.1} = \Lambda(2 \times n_{wvg}) \qquad (5)$$
$$\lambda_{b.r.2} = \Lambda(2 \times n_{SWG}) \qquad (6)$$

To space back-reflections adequately far away from $\lambda_{drop}$, one must simply design the waveguide and SWG to have a sufficient contrast between $n_{SWG}$ and $n_{wvg}$. Assuming no dispersion, the spacing between the dropped wavelength and back-reflected waves, $\Delta\lambda$, can be roughly estimated by

$$\Delta\lambda \sim \Lambda \times (n_{wvg} - n_{SWG}) \qquad (7)$$

## III. DESIGN CONSTRAINTS AND METHOD

To readily implement a SWG filter in GF's 90 nm technology, we designed devices to respect the minimum feature size (MFS) of ~150 nm attainable for bar width in their silicon etch and to function inside the local material stack with a 155 nm thick silicon device layer, buried oxide layer and doped oxide (BPSG) cladding, illustrated in Fig. 2(a). We additionally limited our devices to be constructed using photonic elements available in GF's product design kit (PDK). No new proximity effect calculations were needed for the lithographic steps required to fabricate the SWG filters because GF has used SWG waveguides for spot-size conversion [14]. The SWG filter configuration used here (Fig. 2(b)) is similar to designs previously reported to exhibit good performance metrics in the C-band [15].

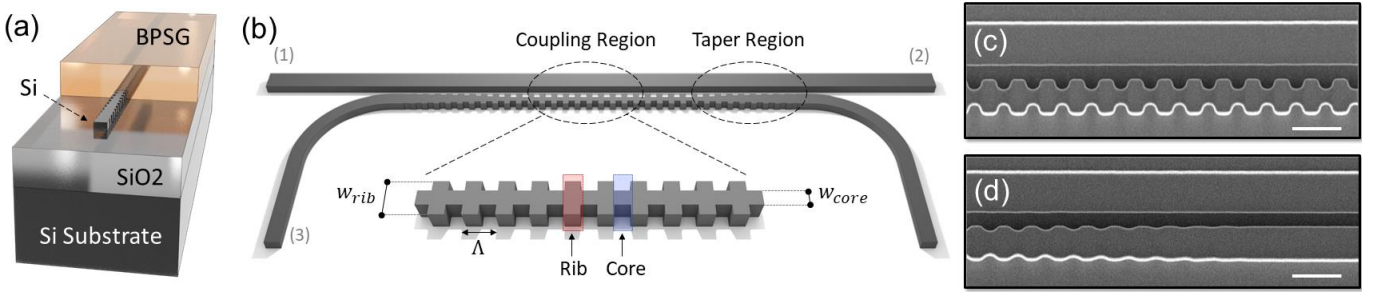

Fig. 2. (a) The material stack used for the SWG filter. (b) Schematic of the full device and port routing with P1, P2 and P3 representing the input, through and output ports, respectively. SEM images of the coupling section and taper region circled in (b) are shown in (c) and (d), respectively. The SEM images clearly show that photolithography can be used to achieve the subwavelength features in the grating. The white scale bars in (c) and (d) are 500 nm long.

A bus waveguide width of 420 nm was chosen to attain a balance between having high effective index and evanescent field extent sufficient for coupling. The designed widths of the periodic high index "rib" regions and low index "core" regions of the SWG, indicated in Fig. 2(b), and the filling fraction of the rib section (FF) were selected to be $w_{rib} = 350$ nm, $w_{core} = 150$ nm, and FF = 50% to ensure sufficient index contrast in the effective indices of the SWG and bus waveguide. The effective index of the SWG mode can be expressed as

$$n_{SWG} = \sqrt{FF \times n_{rib}^2 + (1-FF) \times n_{core}^2}, \qquad (8)$$

where $n_{rib}$ is the effective index of the rib section and $n_{core}$ is the effective index of the core section [15]. These effective indices were calculated in Lumerical MODE Solutions using the cross sections of the rib and core segments and computing the fundamental TE eigenmode for each. Because the bus waveguide was designed to have a higher effective index, we designed the SWG to have a low enough effective index to place reflection-bands outside of the measurement window while still supporting a guided mode. After the calculation of $n_{wvg}$ and $n_{SWG}$, we chose $\Lambda = 330$ nm to filter light in the O-band while respecting the usable MFS. Lumerical FDTD was used to simulate devices and fine-tune drop position after initial calculations using MODE Solutions data. A small taper region (indicated in Fig. 2(b)) was implemented to reduce reflections at the interface of the SWG and the output waveguide. A coupling gap of 160 nm between the waveguide and SWG was chosen. Altering the coupling gap size enables tuning of the drop channel width [15].

## IV. EXPERIMENTAL RESULTS

Filter designs were fabricated in GF's 300 mm foundry in Fishkill, NY (Fab 10). The overall filter footprint was ~0.002 mm$^2$. Scanning electron microscope (SEM) images of the SWG coupling and taper regions, indicated in Fig. 2(b) are shown in Fig. 2(c) and (d), respectively. With the proximity correction implemented in the 90 nm photonics technology, the MFS necessary for the SWG filter (~150 nm) can be readily obtained with photolithography.

After fabrication, devices were characterized at GF's modeling lab in Burlington, VT (Fab 9). As many photonic devices utilize heaters for active tuning, we tested not only the wavelength response of the filter, but also the dependence on filter location and shape with respect to wafer temperature. The ports used in the experiment are indicated in Fig. 1 and Fig. 2(b): (P1) is the input port, (P2) is the through port and (P3) is the drop port used for results in Fig. 3.

The transmission spectra for the through and drop ports are shown in Fig. 3(a), demonstrating an extinction ratio in the through port of ~30 dB for the dropped channel, an insertion loss of ~1.2 dB of the channel in the drop port and a 3 dB channel width of ~5 nm with a flat-top response. The flat-top characteristic of this platform is due to a distribution of power drop rates centered around the drop wavelength specified by (4) and a sufficiently long coupling region to saturate this response in the total dropped power.

The thermal response of the drop port and through port are shown in Fig. 3(b,c), respectively. The channel center red-shifts by ~4 nm as the wafer chuck temperature increases from 25℃ to 85℃. As this temperature was monitored on the wafer chuck, it is possible the local device temperature is lower than the chuck set-point, suggesting that integrated heaters could match or exceed the thermal tuning capabilities reported in Fig 3(b,c).

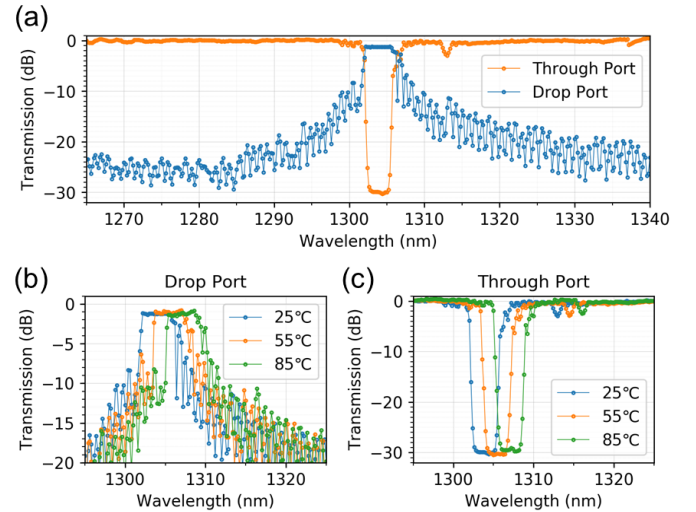

Fig. 3. (a) Transmission spectra of through and drop ports of fabricated device. Thermal response of the (b) drop port and (c) through port. The data clearly shows shape preservation during thermal tuning and a maintained low loss of ~1.2 dB in the drop port.

## V. DISCUSSION OF RESULTS

The SWG filter device realizes low loss (-1.2 dB) in the drop channel and high extinction ratio (-30 dB) in the through port with no increase in loss in the through port compared to a reference waveguide. These results suggest the SWG filter platform could enable a modular design where channels could be independently added, adjusted or removed from a WDM device. The compact footprint of ~0.002 mm$^2$ for the single filter could enable WDM MUX and DEMUX devices with extremely low footprint compared to interferometric platforms.

Additionally, as the drop position solely depends on Λ, $n_{SWG}$ and $n_{wvg}$, the position can be shifted with a single heat source without shape distortion. While many interferometric filters require tuning of all channels simultaneously via multiple heaters, the results here suggest SWG filters require only a single heat source to tune each channel independently, with the tradeoff being that heaters can only red-shift the drop port.

While the filter shown in this work demonstrates desirable performance metrics in loss, channel shape, extinction ratio, and bandwidth, there are design modifications that could improve performance. The proximity of the two straight waveguide segments near the SWG taper (Fig 2(d)) results in a small reflection peak at ~1313 nm in the through port, as shown in Fig. 3(a). Additionally, the abrupt coupling of power to the SWG results in side-band fringes around the drop channel. Prior work reported that the side-band fringes can be reduced significantly with gradual coupling, which can be accomplished by replacing the straight bus waveguide with multiple S-bends [16]. The switch to S-bends could also eliminate coupling between the straight waveguide sections by placing them farther away, removing the reflection peak at ~1313 nm.

## VI. Conclusion

As datacenters have moved from electrical to optical interconnects, the cost of increasing bandwidth is closely tied to the cost of optical transceivers built to accommodate WDM. To mitigate this cost and address the scale of datacenters, integrated photonic solutions are needed to enable WDM in optical interconnects. Toward this goal, we demonstrate the manufacturability and performance of SWG filters in GF's 90 nm monolithic photonics technology. The device, which was designed using GF's PDK and manufactured at Fab 10, demonstrates low channel loss of -1.2 dB, high extinction ratio of -30 dB, 3 dB channel width of ~5 nm, a small footprint of ~0.002 mm$^2$ and a MFS of ~150 nm. Single-source thermal tuning of ~4 nm was achieved in the drop channel without shape distortion. This work demonstrates manufacturability of SWG filters in a monolithic technology and continues the push towards translating high-performance photonics from academia to industry.